\documentclass[11pt]{article}
\usepackage[total={6in, 8in}]{geometry}
\usepackage{graphicx} 
\usepackage{array}
\usepackage{url}
\usepackage{xcolor}
\usepackage{algorithm}
\usepackage{algpseudocode}
\usepackage{amsmath}
\usepackage{comment}
\usepackage{multirow}
\usepackage{hyperref}
\usepackage{url}

\title{Generating heterogeneous data on gene trees}
\author{Mart\'i Cortada Garcia, Adri\`a Di\'eguez Moscard\'o, Marta Casanellas }
\date{}

\begin{document}

\maketitle

\begin{abstract} We introduce GenPhylo, a Python module that simulates genetic data along a phylogeny avoiding the restriction of continuous-time Markov processes.
GenPhylo uses directly a general Markov model and therefore naturally incorporates heterogeneity across lineages. We solve the challenge of generating transition matrices with a pre-given expected number of substitutions (the branch length information) by providing an algorithm that can be incorporated in other simulation software.
\end{abstract}

\section{Introduction}

When testing or training phylogenetic reconstruction methods, it is necessary to have synthetic data simulated by running a Markov process of state substitution on a phylogenetic (gene) tree.
Usually, gene trees trees that represent the evolution of DNA sequences have branch lengths accounting for the expected number of nucleotide substitutions per site.
When the Markov process of nucleotide substitution is assumed to be time-continuous,
generating transition matrices with a pre-given number of substitutions is straightforward and it is what most phylogenetic software does \cite{IQtree, Rambaut1997,asymmetree,pyvolve,genphylodata,symphy}. However, assuming that the process is time-continuous might be very restrictive: less than 4\% of Diagonal Largest in Column (DLC) matrices of size 4 are of continuous-type, see \cite{CFR22} (see also \cite{chang1996} for the importance of DLC in identifiability of substitution parameters). 
Another drawback arises when restricting to continuous-time processes: the simulation of heterogeneous rates across lineages has to be handled via some probabilistic model.

When one avoids the time-continuous assumption and considers a general Markov process, one has to face the difficult task of generating transition matrices with a pre-given number of substitutions. This was addressed in \cite{GenNonh} for very simple models of nucleotide substitution and a partial solution was proposed for the general Markov model (only working for uniform distribution of nucleotides).  

In this paper we present a way of generating a random transition matrix $M$ for an edge $e:u\longrightarrow v$ given an expected number $b$ of substitutions per site and  distribution $\pi$ at $u$. According to Lake \cite{lake1994}, $b$ can be approximated by
\begin{equation}\label{eq:det}
b = -\frac{1}{4}\ln \frac{\sqrt{\det D_\pi}\det M}{\sqrt{\det D_{\pi'}}},
\end{equation}
 where $\pi'=\pi M$ and $D_\pi$ and $D_{\pi'}$ are the diagonal matrices with the corresponding node distributions.

We have implemented this method for DNA and in the software \texttt{GenPhylo} we use it to provide multiple sequence alignments evolving on a phylogenetic tree with given branch lengths. Our algorithm in python can also be imported to be used in larger simulation software such as \cite{pyvolve}.

\section{Design and implementation}

\subsection{Generating DLC Markov matrices with a prefixed number of substitutions}

Consider and edge between two nodes $u \longrightarrow v$. For a given distribution $\pi$ at  node $u$ and for a given branch length $b$ we want to generate a Markov matrix $M$ such that $\pi'=\pi M$ (the distribution at node $v$) and $M$ satisfy \eqref{eq:det}.
As we need to control both the determinant of $M$ and the distribution $\pi M$, we produce $M$ as the product of two Markov matrices, $M=M_1M_2$: $M_1$ will be a random Markov matrix satisfying mild constrains which will give the final distribution $\pi'=\pi M_1$, and $M_2$ will be a time-reversible Markov matrix with stationary distribution $\pi'$ and whose determinant is adjusted such that \eqref{eq:det} is satisfied.
As the determinant of a Markov matrix is bounded above by one, one has to  carefully check that the procedure can be executed.  The full process is described in Algorithms \ref{alg:M1} and \ref{alg:M2} and further explained below.

We use a Dirichlet distribution $Dir(\alpha_1,\alpha_2,\alpha_3,\alpha_4)$ to generate the rows of random Markov matrices. One has to be careful when choosing the parameters for this distribution: on the one hand we want to generate DLC matrices  but on the other hand we need \eqref{eq:det} to hold. To this end, for each row $i$, parameters $\alpha_j$ are set to 1  if $j\neq i$ and parameter $\alpha_i$ is given as a function
$\frac{ke^{-b}}{\sqrt{b}}$ where $k$ is aimed to be the smallest value for which a matrix sampled from this Dirichlet distribution satisfies the desired determinant bounds. These parameters for the Dirichlet distribution in each row are referred to as $Dirichlet(b)$ in Algorithms \ref{alg:M1} and \ref{alg:M2}.

In Algorithm \ref{alg:M2}, we first produce a random time-reversible matrix $M_2$ with stationary distribution $\pi'$ using the Metropolis-Hastings method \cite{dunnMonteCarlo}. Then we seek for a new time-reversible matrix $\tilde{M}_2$ (with same stationary distribution $\pi'$) such that its determinant equals
\begin{equation}\label{eq:d2}
   d_2 = \frac{e^{-4b}\sqrt{\det D_{\pi'}}}{\det (M_1)\sqrt{\det D_{\pi}}}.
\end{equation}
To do so, we consider matrices $\tilde{M}_2=(1-a)M_2+aI$ and determine the value of $a$ by imposing $\det\tilde{M}_2 = d_2$.
   %
    With this process we obtain a Markov matrix with stationary distribution $\pi'$ and with determinant $d_2$.

We check whether the final matrix $M=M_1M_2$ is DLC and we repeat the process otherwise (we set a maximum of 5 repetitions as in most cases the algorithm already produces DLC matrices, even when using long branches as of 1.2).

\begin{algorithm}
\caption{Algorithm to compute $M_1$}\label{alg:M1}
\begin{algorithmic}
\renewcommand{\algorithmicrequire}{\textbf{Input:}}
\renewcommand{\algorithmicensure}{\textbf{Output:}}
\Require node distribution $\pi$, branch length $b$
\While{number of iterations $<50$}
\State $M_1 \gets \textit{Dirichlet}(b)$ 
\State $\pi'\gets\pi M_1$
\If{$\frac{\exp(-4b)\sqrt{\det(D_{\pi'})}}{\sqrt{\det(D_\pi)}}<1$}
\If{$\det(M_1) > \frac{\exp(-4b)\sqrt{\det(D_{\pi'})}}{\sqrt{\det(D_\pi)}}$}
\State \textbf{break} and \textbf{return} $M_1$ 
\EndIf
\Else{}
\State \textbf{repeat}
\EndIf
\EndWhile
\Ensure $M_1$
\end{algorithmic}
\end{algorithm}

\begin{algorithm}[htb]
\caption{Algorithm to compute $M_2$ using Metropolis-Hastings}\label{alg:M2}
\begin{algorithmic}
\renewcommand{\algorithmicrequire}{\textbf{Input:}}
\renewcommand{\algorithmicensure}{\textbf{Output:}}
\Require distribution $\pi_1$, $M_1$, and branch length $l$
\State $P\gets\textit{Dirichlet}(b)$ 
\While{number of iterations $<50$}
\State \textbf{repeat} $(M_{2})_{i,j}\gets \left\{\begin{array}{lc}
       P(i,j)\alpha(i,j) & \text{if } i\neq j\\
       P(i,i)+\sum_{k\neq i}P(i,k)(1-\alpha(i,k)) & \text{otherwise}
         \end{array}\right.$
\State \textbf{until} stationary distribution of $M_2$ approximates  $\pi_1$.
\EndWhile
\State \textbf{then}
\State Find $a \in(0,1)$ root of polynomial s.t. $\tilde{M_2}=(1-a)M_2+aI$ satisfies \eqref{eq:d2} 
\State $M_2=\tilde{M_2}$
\Ensure $M_2$
\end{algorithmic}
\end{algorithm}

\subsection{Sampling sequences from the Markov process on a phylogenetic tree}

Given a Newick tree with branch lengths, we root the tree at an interior node and assign to it a distribution $\pi$ from $Dir(1,1,1,1)$ (unless the distribution is specified by the user). Then for each edge incident to the root $r$ we compute a Markov matrix with the desired branch length as indicated in the previous section. We compute the distribution at the adjacent nodes to the root and repeat the process until a Markov matrix $M^e$ is assigned to each edge $e$.

We generate a sequence of a specific length $L$ at the root $r$ of the tree by sampling from the distribution at the root. Then, for each edge $e$ directed from $r$ and for each site in the sequence we read the nucleotide and we generate the nucleotide at the child node by sampling from the distribution at the corresponding row of the Markov matrix $M^e$. We iterate this process throughout the tree from deeper to outer nodes.


\subsection{Implementation and execution time}

These methods have been implemented in Python3 in our module \texttt{GenPhylo} \cite{Package}, which has been released in the Python Package Index (PyPI) and can be installed using \texttt{pip install GenPhylo}. Additionally, we also offer the option to run the simulator without installing the package by simply cloning our repository \cite{GenPhylo} and running the \texttt{GenPhylo.py} script from the terminal with the appropriate parameters.


In Table \ref{table:time} we provide the running time of our algorithm. Note that our algorithm encompasses a rich level of heterogeneity at the expense of speed.



\begin{table}[h!]
\centering
\begin{tabular}{|c|c|*{2}{>{\centering\arraybackslash} p{4em}}|}
\hline
\textbf{Leaves} & \textbf{Branch length} & \multicolumn{2}{|c|}{\textbf{Alignment length}}
\\ \cline{3-4}
& & 1000 &10000 \\
\hline
\multirow{2}{*}{4}  & {0.1}  & 1.20  & 1.51 \\ \cline{2-4}
                    & {0.5}  & 1.33 & 1.76 \\ \hline
\multirow{2}{*}{8}  & {0.1} &  2.22 & 3.06 \\ \cline{2-4}
                    & {0.5} & 2.64  & 3.53 \\ \hline
\multirow{2}{*}{16}  & {0.1}  & 4.15  & 6.17 \\ \cline{2-4}
                    & {0.5}   & 5.09  & 7.11 \\ \hline
\end{tabular}
\caption{\label{table:time} Average execution time for generating $100$ alignments on balanced trees with the given branch lengths (same at all branches) and alignment lengths. Computations run on a M1 Mac with an ARM-based processor.}
\end{table}


%

%
%
%

\section*{Author's contributions}
Last author conceived the idea and the first two authors equally contributed to develop and implement it; all authors wrote and revised the manuscript.

\section*{Acknowledgments}
We thank Jes\'us Fern\'andez-S\'anchez for useful discussions on this topic. MC was partially supported by grants PID2019-103849GB-I00 and PID2023-146936NB-I00 funded by MICIU/AEI/
10.13039/501100011033 and Mar\'{\i}a de Maeztu Program for Centers and Units of Excellence in R\&D (project CEX2020-001084-M) and the AGAUR project 2021 SGR 00603 Geometry of Manifolds and Applications, GEOMVAP.

\bibliographystyle{apalike}


\end{document}